\documentclass[prd,preprint,nofootinbib,nobibnotes]{revtex4}
\usepackage{epsf,epsfig}

\newcommand{\be}{\begin{equation}}
\newcommand{\ee}{\end{equation}}
\newcommand{\beq}{\begin{equation}}
\newcommand{\eeq}{\end{equation}}
\newcommand{\ben}{\begin{displaymath}}
\newcommand{\een}{\end{displaymath}}
\newcommand{\beqa}{\begin{eqnarray}}
\newcommand{\eeqa}{\end{eqnarray}}
\newcommand{\bea}{\begin{eqnarray}}
\newcommand{\eea}{\end{eqnarray}}
\newcommand{\bean}{\begin{eqnarray*}}
\newcommand{\eean}{\end{eqnarray*}}
\newcommand{\ba}{\begin{array}}
\newcommand{\ea}{\end{array}}
\newcommand{\bi}{\begin{itemize}}
\newcommand{\ei}{\end{itemize}}

\newcommand{\eg}{{\it e.g.,\,}}

\newcommand{\reef}[1]{(\ref{#1})}

\begin{document}

\begin{flushright}
hep-th/0607023
\end{flushright}

\vskip 1.2in

\title{Microstates of a Neutral Black Hole in M Theory}
\author{Roberto Emparan$^{a,b}$ and Gary T.~Horowitz$^{c}$ \\}
\address{ 
$^a$\,Instituci\'o Catalana de Recerca i Estudis
Avan\c cats (ICREA)\\ 
$^b$\,Departament de F{\'\i}sica Fonamental,
Universitat de Barcelona, Diagonal 647, E-08028 Barcelona, Spain \\
$^c\,$Department of Physics, University of California, Santa
Barbara, CA 93106-9530, USA \\
{\small \tt emparan@ub.edu,
gary@physics.ucsb.edu} \\
}
\begin{abstract}
We consider vacuum solutions in M theory of the form of a five-dimensional
Kaluza-Klein black hole cross $T^6$. In a certain limit, these include
the five-dimensional neutral rotating black hole (cross $T^6$). From a IIA
standpoint, these solutions carry D0 and D6 charges. We show that there
is a simple D-brane description which precisely reproduces the
Hawking-Bekenstein entropy in the extremal limit, even though
supersymmetry is completely broken. 
\end{abstract}
\maketitle

\section{Introduction}

An important deficiency in the string theory of black holes is the fact
that the simplest solutions, \eg\  Schwarzschild or Kerr, are
rather different than those whose entropy has been statistically
reproduced. The original black hole studied by Strominger and Vafa 
\cite{SV} was  charged and supersymmetric. 
The charges of a black hole serve as tags that help
identify its microscopic constituents in string theory. In addition, when
the solution is supersymmetric, the phase space of the system is
drastically constrained and subject to powerful non-renormalization
theorems and state-counting techniques. Neutral black holes, however,
carry a minimal set of quantum numbers---mass and angular
momentum--- and so it seems hard to restrict the phase space to a sector
which is  simple enough  to count the microstates.

Nevertheless, we will argue that there exist vacuum black holes in M
theory that can be mapped to well-defined bound states of D-branes in
string theory, and which, in certain limits, become asymptotically flat
black holes. We start with dyonic solutions of five-dimensional
Kaluza-Klein (KK) theory \cite{gw}. Ref.~\cite{itz} showed that in a
certain limit, akin to the decoupling limit in AdS/CFT, these solutions
include the five-dimensional neutral rotating black hole of Myers and
Perry \cite{MP}, which is asymptotically flat. Reversing this procedure,
one can view the KK black hole as the Myers-Perry black hole placed at
the tip of a Taub-NUT geometry. A similar connection between four and
five dimensional black holes has been discussed recently for
supersymmetric solutions with additional charges and self-dual angular
momentum \cite{Gaiotto:2005gf}. Here we are considering the simplest
case of a black hole with generic $J_1$ and $J_2$ and no extra charges.

Taking the product with a flat $T^6$, we obtain a solution to M theory,
whose IIA reduction has D0 and D6 charge. 
Even in the extremal limit, this black hole is not supersymmetric
\cite{KO}, in accord with the absence of supersymmetric bound states of
D0 and D6 branes. There are, however, non-supersymmetric, quadratically
stable, D0-D6 bound states \cite{wati}, and these will serve as a basis
to our microscopic picture. We will provide a simple string
description that exactly reproduces the entropy and mass of the extremal black
hole. 

The entropy of some nonsupersymmetric, extremal black holes has been
reproduced before \cite{nosusy}. However, that required black holes with
four charges (in four dimensions) while we have only two. More
importantly, unlike previous examples our solutions are pure vacuum in
higher dimensions. The entropy of neutral black holes can be understood
in terms of a correspondence principle \cite{Horowitz:1996nw}, but that
does not reproduce the precise coefficient.
Earlier work attempting to obtain statistically the precise entropy of
asymptotically flat neutral black holes by different means include
\cite{argurio,ddbar}. Previous attempts at providing a microscopic
description of D0-D6 black holes include
\cite{sheinblatt,larsen2}.

\section{Kaluza-Klein and Myers-Perry black holes}

We begin by reviewing the KK black holes (for a detailed description,
see \cite{gw}). 
These black holes are characterized (in four dimensions) by their mass
$M$, angular momentum $J$, and electric
and magnetic
charges $Q$
and $P$. They satisfy the inequality
\beq\label{bound}
2G_4M\geq \left(Q^{2/3}+P^{2/3}\right)^{3/2}\,,
\eeq
which, at slow rotation $G_4 J<PQ$, is saturated in the extremal limit
independently of $J$. When
$P=Q$ and $J=0$ the
four-dimensional geometry becomes exactly the same as the
Reissner-Nordstrom black hole. 

In the five-dimensional vacuum solution, let $y$ be the  compact KK
dimension, $y\equiv y+2\pi R$. This circle is fibered over the two-spheres
of spherical symmetry. Since $S^1$ bundles over $S^2$ are labeled by an
integer,  the magnetic
charge must be quantized in terms of the radius $R$. The electric charge
is also quantized, since it corresponds
to momentum in the $y$-direction. More precisely,
\beq\label{nsix}
 \qquad Q=\frac{2G_4 N_0}{R},  \qquad  P=\frac{N_6 R}{4}
\eeq
 for integers $N_0$ and $N_6$ (the reason for this notation will become
clear below).

The five-dimensional interpretation of these solutions is quite
interesting. In the absence of magnetic charge, the horizon has topology
$S^1\times S^2$, where $S^1$ is the KK circle, and the
solution is a black string boosted along $y$. However, the topology
changes when $P\neq 0$. If $N_6=1$, the $y$-circle and spherical $S^2$
combine into a topological $S^3$. In the extremal limit with $Q=0$ and $J=0$, the
solution becomes the KK monopole. The geometry can be
described as a `cigar' fibered on the orbital $S^2$. 

If we add electric charge or energy above extremality, we find a finite
black hole horizon at the tip of the cigar. So magnetically charged
KK black holes are five-dimensional black holes with horizon
topology $S^3$, localized inside a Taub-NUT geometry.
The electric charge does not
correspond to a boost, but rather to \textit{rotation} of the black hole
aligned with the KK circle. A component of the rotation that
is not aligned with the five-dimensional fiber gives rise to
four-dimensional rotation.

If the size of the black hole is much smaller than the
KK radius $R$, then finite-size effects become
negligible and we recover the five-dimensional Myers-Perry black hole,
as explained in \cite{itz}. 
In this limit, the four-dimensional mass is dominated by the mass
of the KK monopole, and
the excitation energy above the KK monopole
is equal to the ADM mass of the five-dimensional black hole.
The angular momenta in five
dimensions are related to the electric
charge and four-dimensional angular momentum as
$J_1+J_2=N_0 N_6^2$ and $J_1-J_2=2J N_6$.

The identification with a five-dimensional black hole is a local one.
Globally, 
the asymptotic spatial geometry is actually the orbifold
$\mathbf{R}^4/\mathbf{Z}_{N_6}$. So only
configurations with $N_6=1$ give rise to globally
asymptotically flat solutions. When $N_6>1$ the black hole sits at the
tip of a conical space.

The entropy of the KK black hole is particularly simple in the
extremal limit \cite{larsen2},
\beq\label{Sbh}
S=\frac{A_{(4)}}{4G_4}=2\pi\sqrt{\frac{P^2Q^2}{G_4^2}-J^2}
=2\pi \sqrt{\frac{N_0^2 N_6^2}{4}-J^2}\,.
\eeq
This is independent of the circle radius $R$, so it also corresponds to
the entropy of the extremal
Myers-Perry black hole after the limit of infinite radius $R$ is taken.
It was noted in \cite{Dhar:1998ip,larsen2} that the entropy depends only
on the integer normalized charges. This is a strong indication that a
microscopic counting of the states is possible.

\section{Microscopic description}

In order to count the microstates of the KK black hole, we  take
the product with $T^6$ (with volume $(2\pi)^6 V_6$) and view it as a
vacuum solution in M theory with the KK circle being the M theory
circle. By the usual relation between M theory and IIA string theory,
$R=g l_s$. The electric and magnetic charges now correspond to D0 and
D6-branes, and $N_0$ and $N_6$ are simply the net number of D-branes.
The quantization condition (\ref{nsix}) can now be written
\be
Q= 2G_4 M_0 N_0\,, \qquad P= 2G_4 M_6 N_6
\ee
where the masses of individual D0 and D6-branes are
\be
M_0 = {1\over g l_s  }\,, \qquad M_6 = {V_6\over g l_s^7 }\,
\ee
and $G_4=g^2 l_s^8/8V_6$. So the ADM mass of our extremal black hole is
\be\label{bhmass}
M= [(M_0 N_0)^{2/3} + (M_6 N_6)^{2/3}]^{3/2}\,.
\ee
To reproduce this mass and the entropy formula (\ref{Sbh}) we will pass
to a T-dual configuration where the microscopic description becomes more
transparent. For simplicity we consider first the case without
four-dimensional rotation, $J=0$, and will discuss $J\neq 0$ near the end.

We first recall the situation for the supersymmetric four-charge
black holes in Type II string theory compactified on $T^6$. There are
many possible choices for the charges, all related by U-duality. For our
purposes, the most useful is in terms of four stacks of D3-branes
\cite{Balasubramanian:1996rx}. Any two stacks intersect over a line, and
all four intersect at a point. The orientation of the first three stacks
can be chosen arbitrarily, but to preserve supersymmetry, the
orientation of the last set of D3-branes is then fixed. We are
interested in the case where the number of branes in each stack is the
same, say $N$. The moduli of the $T^6$ then remain constant, and the
solution reduces to the product of
$T^6$ and extreme Reissner-Nordstrom,
 \be
ds^2 = -\left(1+{r_1\over r}\right)^{-2} dt^2 + \left(1+{r_1\over
r}\right)^2(dr^2 + r^2 d\Omega_2)\,.
\ee
Assuming a square torus with equal size circles and $V_6=(V_3)^2$, the
constant $r_1$ is related to the number of 3-branes $N$ via
\beq
r_1= \frac{g N l_s^4}{2V_3}\,.
\eeq
The ADM mass is
\be
M = {r_1\over G_4} = {4N V_3\over g l_s^4 }
\ee
which is just the mass of the four stacks of $N$ 3-branes wrapped around
the torus, and the black hole entropy is
\be\label{S3branes}
S = {A\over 4G_4} = 2\pi N^2\,.
\ee
Although the explicit counting of states for supersymmetric
four-dimensional black holes is easier to carry out with a different
choice of charges \cite{Maldacena:1996gb}, the fact that it reproduces
(\ref{S3branes}) and is related by U-duality ensures that the D3-branes
also contain precisely the right number of states (at large $N$) to
reproduce the black hole entropy. Furthermore, since the entropy is
independent of the moduli of the torus, it seems clear that the states
are associated with the intersection point of the branes.

\begin{figure}
\begin{center}\leavevmode  %
\epsfxsize=4.5cm 
\epsfbox{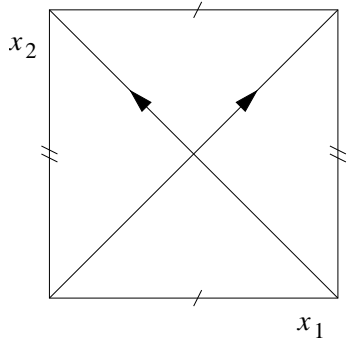}
\end{center}
\caption{\small Branes wrapping the diagonals of a torus. There are two
intersection points, at the origin and at the middle of the square. We
assume that each
intersection contributes a microscopic entropy equal
to that of a supersymmetric intersection of branes.
}
\label{fig:diagonals}
\end{figure}

Bound states of four D0-branes and four D6-branes were described by
Taylor in terms of a gauge theory configuration on the worldvolume on
the 6-brane \cite{wati}. He pointed out that after applying T-duality
along three cycles on the torus, this configuration was equivalent to
four D3-branes in a configuration very similar to the one described
above. However, there are two important differences. The orientations
correspond to broken supersymmetry, and the branes are wrapping the
diagonals of the torus. To be explicit, consider first a square $T^2$ with
coordinates $(x_1,x_2)$. The two diagonals are given by $x_2 = \pm x_1$
which we will call the $+$ and $-$ cycle. If we orient the cycles so
that $x_2$ always increases, then a configuration of two strings
wrapping both diagonals has net winding number two around $x_2$ and zero
winding around $x_1$ (see Fig.~\ref{fig:diagonals}). 
Now view $T^6$ as the product of three $T^2$'s,
with coordinates $(x_1,x_2), (x_3,x_4)$ and $(x_5,x_6)$ respectively.
The 3-branes are all wrapped around one diagonal of each $T^2$ and
oriented so that the even coordinates always increase. So the
configuration can be labelled by specifying which of the diagonals is
wrapped on each $T^2$. If we T-dualize in the $2,4,6$ directions,  the
configuration dual to the four D0-branes and
four D6-branes is 
\be\label{config}
(+++), (+--),(-+-),(--+)
\ee
where the first entry corresponds to the first torus, etc. By
construction, each brane wraps the cycle (246) once, and since each
entry has an even number of minus signs, each brane also wraps the cycle
(135) once. It is easy to check that the net winding about any other
3-cycle (such as (146)) is zero. So even though there are four 3-branes,
the net nonzero charges are just (135) and (246), which is what one
expects after three T-dualities of D0-D6. 

If we replace the single 3-brane around each cycle in (\ref{config}) by
$N$, we obtain a configuration with charge $4N$ around each of the
cycles (135) and (246). This reproduces the ADM mass of the black hole.
After three T-dualities, we get $N_0=N_6=4N$ and hence (\ref{bhmass})
becomes $ M = 2^{3/2} 4N V_3/ g l_s^4$ \footnote{Starting with a symmetric $T^6$
for the intersecting 3-branes and applying three T-dualities results
in a torus with volume $V_6= l_s^6$  (and changes the string coupling to $g l_s^3/V_3$). So the 6-brane and 0-brane have equal
mass.}. This is equal to the mass of the 3-branes since the length of each
leg is $\sqrt 2$ larger than before, so the volume of
each 3-brane is $2^{3/2}$ greater.
The fact that the mass does not saturate a BPS bound is just a
reflection of the fact that the branes are wrapping cycles of larger
volume.

What about the entropy? At first sight there appears to be a
discrepancy. If we make the reasonable assumption that the entropy of
the intersecting 3-branes is unaffected by the change in orientation
and rotation of the branes we would expect $S= 2\pi N^2$ as in
\reef{S3branes}. However, since
$N_0 = 4N$ and $N_6=4N$ (and $J=0$) the black hole entropy \reef{Sbh} is
\be
S_{bh} = 16\pi N^2
\ee
which is larger by a factor of eight. However, rotating the branes
increases the number of intersection points \footnote{We thank Juan
Maldacena for suggesting this.}. On a $T^2$, the diagonals have two
intersection points (see Fig.~\ref{fig:diagonals}). Since the branes
have two intersection
points on each of the three $T^2$'s, there are a total of eight
intersection points. 
 The total Hilbert space is a
tensor product of the states at each intersection point and hence 
\beq
S_{branes}=8\times 2\pi N^2\,.
\eeq
Thus, a simple weak coupling calculation reproduces
the black hole entropy exactly.

It is easy to generalize this to the case of unequal charges (in terms
of gauge fields on the 6-brane, this was done in
\cite{Dhar:1998ip,larsen2}). The configuration of branes is again given
by (\ref{config}) where $\pm$ now refer to more general cycles than just
the diagonal. Let $\pm$ denote the cycles $x_2= \pm k x_1/l$ for
relatively prime integers $k,l$ (and similar cycles on the other two
$T^2$'s with the same integers $k,l$). The configuration of branes
(\ref{config}) now has charge $4k^3$ along (246) and charge $4l^3$ along
(135). The mass of each brane is now $(k^2 +l^2)^{3/2}$ larger just from
the increase in area of the three-cycle being wrapped. This agrees with
the ADM mass since with $N$ branes wrapped around each of the cycles,
$N_0=4 k^3 N$, $N_6=4 l^3N$ so (\ref{bhmass}) yields
\be
 M_{bh} = \frac{4N(k^2 + l^2)^{3/2}\;V_3}{g l_s^4 } = M_{branes}\,.
  \ee
In retrospect, the presence of $3/2$ in the exponent  of the black hole mass
is an indication of
a microscopic description in terms of 3-branes.

\begin{figure}
\begin{center}\leavevmode %
\epsfxsize=11cm 
\epsfbox{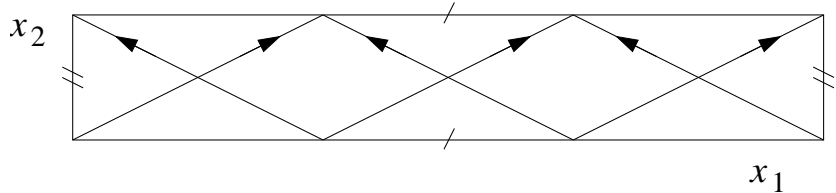} 
\end{center} 
\caption{\small Generalization to unequal 
charges and non-trivial moduli. The branes wrap a rational direction 
$k/l$ of the torus (in the figure, $k=3$, $l=1$), so there are $2kl$ 
intersection points on each $T^2$. In the limit to the 
five-dimensional Myers-Perry black hole, the torus shrinks along 
$x_{2}$. } 
\label{fig:unequal} 
\end{figure}

The entropy also comes out exactly right since the $+$ and $-$ cycles
now have $2kl$ intersection points on each $T^2$ (see
Fig.~\ref{fig:unequal}). So the
collection of
3-branes has a total of $(2kl)^3$ intersection points. The entropy is
thus
\bea\label{Sexact}
S_{branes} &=& (2kl)^3 \times 2\pi N^2 = \pi (4Nk^3)(4N l^3) 
\nonumber\\
&=&\pi N_0 N_6= S_{bh}\,.
\eea

The limiting case of a Myers-Perry black hole described above 
requires only a slight generalization (see Fig.~\ref{fig:unequal}). 
We want to take $R\rightarrow \infty$ keeping $G_5$, $G_{11}$, $N_0$ and
$N_6$ fixed. In the D0-D6 frame, this corresponds to taking
$g\rightarrow \infty$ keeping $V_6$ fixed in Planck units. Since the
eleven-dimensional Planck length $l_p$ is given by $l_p= g^{1/3} l_s$,
when we T-dualize along a direction, the new circle has length $\tilde L
\sim l_s^2/L \sim g^{-2/3}$. Thus, after T-duality in the $2,4,6$
directions, the size of these three circles goes to zero in the limit.
(The IIB string coupling remains finite since $\tilde g \sim g l_s^3/L^3
\sim l_p^3/L^3$). We again obtain 3-branes wrapping the cycles
(\ref{config}), but they become essentially parallel as we approach the
Myers-Perry black hole, all wrapping the (135) cycle with a positive
orientation. Since the entropy is moduli-independent, the equality
between statistical and black hole entropies holds as in \reef{Sexact}. 

Finally, to allow for $J\neq 0$ we assume that $J$ is evenly distributed
among the $(2kl)^3$ intersections of 3-branes so each one carries
angular momentum $J_0 = J/(2kl)^3$. In the $(0,4)$ theory that describes
the four-charge system, to account for $J_0$ we align the polarization of
$J_0^2/N^3$ fermionic left-moving excitations (out of $N$) while the
right-movers remain unexcited. The entropy is then $2\pi \sqrt{N^3(N -
J_0^2/N^3)}$. (This is the nonsupersymmetric analog of \cite{HLM}).
Assuming that this applies to each intersection, the mass formula is not
modified
but the total microscopic entropy becomes
\beq
S_{branes}=(2kl)^3 \times 2\pi \sqrt{N^4 - J_0^2}
 =2\pi \sqrt{\frac{N_0^2 N_6^2}{4} - J^2}\,.
\eeq
This reproduces \reef{Sbh} and hence also the entropy of the extremal
rotating Myers-Perry black holes with generic angular momenta.

\section{Discussion}

It seems remarkable that a simple system of D-branes is able to
reproduce the mass and entropy of a vacuum black hole which is far from
being supersymmetric. It is not clear to us why this is working so well,
but it hints at further simplifications that might be possible for
neutral black holes. In particular, it is intriguing that we need to use
four different sets of branes, even though (from the IIA standpoint)
there are only two charges. This is reminiscent of
earlier suggestions that neutral black holes should be viewed as
collections of branes and anti-branes \cite{Horowitz:1996ay,ddbar}.
There is a mysterious duality invariant formula which reproduces the
entropy of all nonextremal black holes (including Schwarzschild) in
terms of branes and antibranes. It is not yet clear how to derive this
in string theory, but the above construction seems a
step in the right direction.

Various other open questions remain: (1) We have only considered extreme
KK black holes. Can one count the microstates of near-extremal
solutions? This would bring us a little closer toward understanding
Schwarzschild. (2) Can one reproduce the entropy of KK black holes with
$N_6=1$? The constructions above seem to require that $N_6\ge 4$
(actually $N_6\gg 4$, since assigning an entropy $2\pi N^2$ to each
intersection is justified only for large $N$). (3) Can one replace $T^6$
with general Calabi-Yau spaces and still count the entropy? Since mirror
symmetry can be viewed as T-duality on a $T^3$, an initial collection of
D0 and D6-branes goes over to a collection of D3-branes under this
symmetry. This suggests that the above construction may have a natural
generalization. (4) Rather than working with D3-branes, can one
understand the entropy of the above black holes directly in M theory (in
terms of gravitons and perhaps branes) or in terms of D6-branes with
flux corresponding to D0-branes? This latter case corresponds to
counting the number of instantons (of a certain type) in six-dimensional Yang-Mills
theory.

\begin{acknowledgments}
We thank the KITP, Santa Barbara for the stimulating program ``Scanning
New Horizons: GR Beyond 4 Dimensions" where this work was begun. We also
thank R.~Kallosh, F.~Larsen, J.~Polchinski, S.~Trivedi, and especially
J.~Maldacena for discussions. GH thanks the IAS Princeton for their
hospitality. RE was supported in part by DURSI 2005 SGR 00082, CICYT FPA
2004-04582-C02-02 and EC FP6 program MRTN-CT-2004-005104. GH was
supported in part by NSF grants PHY-0244764 and PHY-0555669.
\end{acknowledgments}


\end{document}